\documentclass{article}
\usepackage{arxiv}
\usepackage{balance}
\usepackage{enumitem}
\usepackage{amsmath,amssymb,amsfonts}
\usepackage{algorithmic}
\usepackage[table]{xcolor}
\usepackage{soul}
\usepackage{graphicx}
\usepackage{textcomp}
\usepackage{siunitx} 
\usepackage{bmpsize}
\usepackage{url}
\usepackage{lipsum}
\usepackage{tabularx}
\usepackage{tikz}
\usetikzlibrary{positioning, shapes.multipart, arrows.meta}
\usepackage{booktabs}
\usepackage{subcaption}
\usepackage{fancybox}
\usepackage{multirow}
\usepackage{mdframed}
\usepackage{tcolorbox}
\usepackage{cleveref}
\def\BibTeX{{\rm B\kern-.05em{\sc i\kern-.025em b}\kern-.08em
    T\kern-.1667em\lower.7ex\hbox{E}\kern-.125emX}}

\def\sbomfiles{620}
\def\totalGHrepositories{26,625}
\def\uniquerepositories{152}
\def\totalGHandSGrepositories{26,823}

\def\deduplicatedleftSGrepos{198}

\def\dependencies{181,283}
\def\uniquedependencies{25,430}

\newmdenv[
    backgroundcolor=gray!10,
    frametitlebackgroundcolor=darkgray!20,
    frametitlefont={\color{black}\bfseries},
    frametitlealignment=\hspace{0pt}
]{rq:answer}

\newcommand{\takeaway}[2]{
\begin{mdframed}[backgroundcolor=gray!10,linewidth=2pt, leftline=true, topline=false, bottomline=false, rightline=false]
{\bfseries Takeaway #1}: #2
\end{mdframed}
}

\begin{document}

\title{Policy-driven Software Bill of Materials on GitHub: An Empirical Study}
\author{Oleksii Novikov \\
    Blekinge Institute of Technology\\
    Karlskrona, Sweden \\
    \texttt{oleksii.novikov@bth.se} \And
Davide Fucci \\
    Blekinge Institute of Technology\\
    Karlskrona, Sweden \\
    \texttt{davide.fucci@bth.se} \And
Oleksandr Adamov  \\
    Blekinge Institute of Technology\\
    Karlskrona, Sweden \\
    \texttt{oleksandr.adamov@bth.se} \And
Daniel Mendez \\
    Blekinge Institute of Technology\\
    Karlskrona, Sweden \\
    \texttt{daniel.mendez@bth.se} 
}
\maketitle
\begin{abstract}
\textit{Background}. The Software Bill of Materials (SBOM) is a machine-readable list of all the software dependencies included in a software.
SBOM emerged as way to assist securing the software supply chain.
However, despite mandates from governments to use SBOM, research on this artifact is still in its early stages.
\textit{Aims}. We want to understand the current state of SBOM in open-source projects, focusing specifically on \textit{policy-driven} SBOMs---i.e., SBOM created to achieve security goals, such as enhancing project transparency and ensuring compliance, rather than being used as fixtures for tools or artificially generated for benchmarking or academic research purposes.
\textit{Method}. We performed a mining software repository study to collect and carefully select \sbomfiles{} SBOM files hosted on GitHub. 
We analyzed the information reported in policy-driven SBOMs and the vulnerabilities associated with the declared dependencies by means of descriptive statistics. 
\textit{Results}. We show that only 0.56\% of popular GitHub repositories contain policy-driven SBOM. The declared dependencies contain 2,202 unique vulnerabilities, while 22\% of them do not report licensing information.
\textit{Conclusion}. Our findings provide insights for SBOM usage to support security assessment and licensing.

\end{abstract}

\section{Introduction}
A Software Bill of Materials (SBOM) lists all components that go into a piece of software, making its supply chains more transparent for those who use, make, buy, or regulate it.
Creating and maintaining an SBOM is a security practice offering a mechanism to quickly check if software and its dependencies are affected once vulnerabilities are disclosed.
The Log4J incident~\cite{feng2022defense} demonstrated the importance of an SBOM.
When the Log4Shell vulnerability (CVE-2021-44228) was discovered in the Apache Log4J library, organizations with an SBOM could quickly analyze their dependency tree and identify whether their software included the vulnerable dependency version.
A comprehensive SBOM would have allowed them to take immediate action to mitigate risk, patch the vulnerability, or apply the necessary workarounds.
The rapid increase in software supply chain attacks in the last few years (e.g., SolarWind~\cite{wolff2021navigating}, event-stream~\cite{arvanitis2022systematic}, Log4J~\cite{everson2022log4shell}) resulted in the US Government mandating the use of SBOM for their suppliers
 and with similar actions being taken in the context of the European Union Cyber Resilience Act (EU CRA)\footnote{\url{http://data.europa.eu/eli/reg/2024/2847/oj}}.
Regarding licensing, SBOM helps organizations comply with open-source license requirements, identify risks due to restrictive licensing (e.g., AGPLv3.0), and support auditing automation (e.g., verifying that software components meet corporate licensing policies).   
Although SBOMs are crucial for managing dependencies, vulnerabilities, risks, and licenses~\cite{Zahan.20236zj}---and despite a growing interest~\cite{hendrick2022state}---industrial adoption is limited~\cite{kloeg2024charting,stalnaker2024boms}.

At the same time, scientific research on SBOM is still in its infancy.
One significant barrier is that their generation varies across the software development lifecycle, without a standard approach to when and how to create and update them~\cite{Xia2023}.
The value of SBOMs is further questioned by the tooling used to generate them, lack of interoperability among formats (e.g., SPDX, CycloneDX, SWID)~\cite{Zahan.20236zj}, and the varying quality, accuracy, and completeness of their contents~\cite{Torres-Arias2023a,Mirakhorli24}.  
In this respect, \textit{most investigations rely on synthetic ad-hoc SBOMs}---i.e., generated by the researchers themselves from a source code repository or other sources (i.e., a Docker image)---rather than created by practitioners for a practical purpose~\cite{o2024assessing}.
This reliance on artificial SBOM limits the ability to draw realistic conclusions about their effectiveness as they often fail to capture the context of real-world software projects, resulting in studies that may not reflect practical issues.
On the other hand, SBOMs created intentionally by practitioners as part of a policy are needed to understand their utility, better address issues, and develop best practices for their creation, consumption, and maintenance.
In this work, we specifically focus on such type of SBOM files, which we call \textit{policy-driven SBOM}. 
Although several policies can drive the creation of SBOM, in this work we focus on polices related to security risk assessment, supply-chain transparency, and compliance.
Accordingly, our study is driven by the following overarching goal.

\begin{mdframed}[backgroundcolor=gray!10, leftline=false, topline=false, bottomline=false, rightline=false]
Understand the characteristics of policy-driven SBOM files found in open-source projects.
\end{mdframed}
To that end, we carefully collect and analyze SBOM files mined from GitHub repositories.
Moreover, we associate known vulnerabilities to the dependencies specified in the SBOM files we collected. 
We show how SBOM is currently used in open-source projects and can inform further work supporting vulnerability and license management.
The contribution of this paper is threefold:
\begin{itemize}
    \item It defines a simple taxonomy to identify policy-driven SBOM instead of synthetic ones.
    \item It proposes a methodology to mine policy-driven SBOM files from open-source software repositories.
    \item It provides a dataset\footnote{\url{https://github.com/AleX04Nov/sbom_scanning}} including \sbomfiles{} policy-driven SBOM files and an analysis of the listed \uniquedependencies{} dependencies.
\end{itemize}

This paper is organized as follows. 
\Cref{sec:background} introduces the main terminology and ideas behind our work and reviews related literature investigating SBOM, whereas \Cref{sec:methodology} presents the research question driving this work and the methodology we followed to gather policy-oriented SBOMs, their quality, and associated vulnerabilities. 
\Cref{sec:results} provides an overview of the results while \Cref{sec:directions} discusses their implication.
\Cref{sec:conclusion} concludes the paper.

\section{Background and Related Work}
\label{sec:background}
This section establishes relevant concepts and surveys the related literature.

\subsection{Background}
Following the work of Torres-Arias et al.~\cite{Torres-Arias2023a}, SBOM files can be categorized into two types, \textit{In-the-Lab} and \textit{In-the-Wild}. 
In-the-Lab SBOMs are usually generated for research purposes, typically to compare SBOM generation tools.
For example, Balliu et al.~\cite{balliu2023} created SBOM files to highlight the issues with Java-related tools for SBOM generation.
SBOM files can be artificially created for comparing vulnerability detection approaches or tools~\cite{balliu2023,YU2024,Rabbi2024}.
Moreover, researchers compare SBOM by generating them using the same tool but in different formats, such as SPDX and CycloneDX~\cite{Mirakhorli24}.

In-the-Wild SBOMs are \textit{not} generated for research purposes but rather intending to support the following use cases: dependency management, security scanning, licensing, testing of tools, and tracking 3rd party dependencies.

\textit{Dependency management} supports developers and other stakeholders to understand the components that are eventually included in a software.
\textit{Security scanning} supports developers in finding vulnerabilities in their dependencies, usually providing SBOM as input to automated vulnerability scanning tools.
\textit{Licensing} supports developers and other stakeholders in dealing with issues relating to, for example, conflicting usage terms between dependencies included in a software. 
However, in the wild, some SBOM files report licensing information for the project itself rather than for its dependencies. We refer to these as \textit{Self-License} SBOM files.
These SBOMs substitute \texttt{LICENSE.md} files commonly found in open-source repositories~\cite{vendome2017license} and do not report any components or dependencies information.
SBOM files can also be used as \textit{Test Data} (e.g., test data) for tools that need, for example, to parse SBOM and for testing such tools.

Finally, \textit{3rd party SBOM files} are generated by dependency management tools based on external packages---i.e., they represent transitive dependencies. 
The SBOM of a project  should only include its direct dependencies (e.g., as required by the EU CRA). Transitive dependencies from third parties should be accessed by traversing the dependency tree formed by each individual SBOM.

\begin{figure*}
\centering
    \includegraphics[width=\textwidth, trim={0 0.4 0 0.4cm},clip]{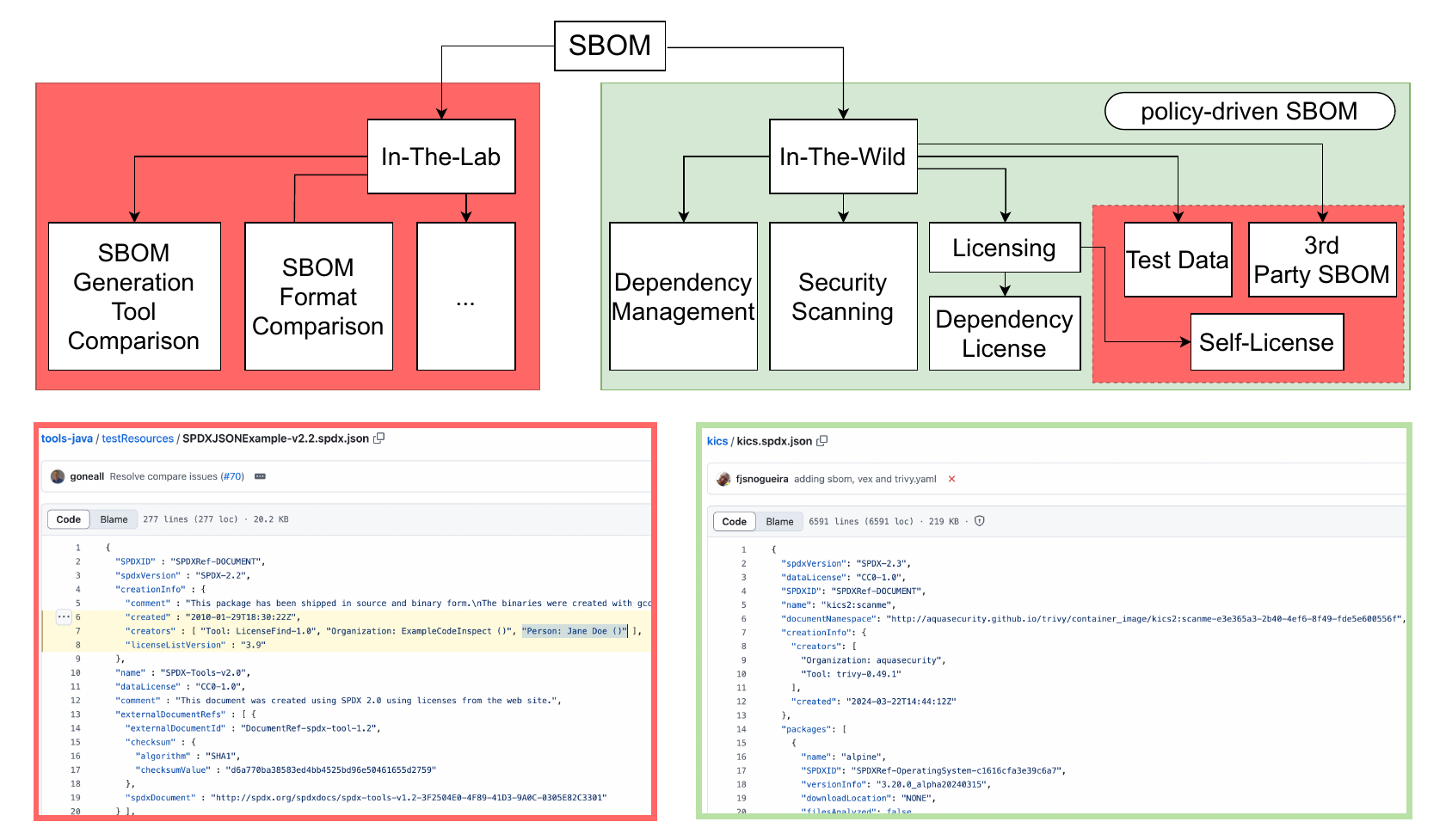}
    \caption{Taxonomy of SBOM types (green considered in our study, red excluded) with examples.}
    \label{fig:examples}
\end{figure*}

In our research, we focus on the practical uses of SBOM files. 
Therefore, we do not consider \textit{In-the-Lab} as well as \textit{Test Data}, \textit{3rd party SBOM}, and \textit{Self-License} SBOM.
Our focus remains on use cases that have a direct impact on a project in terms of security and compliance---i.e., \textit{Dependency Management}, \textit{Security scan}, and \textit{Licensing}---independently of the nomenclature suggested by the Cybersecurity and Infrastructure Security Agency~\cite{CISA}.
In other words, we do not distinguish between the different software lifecycle phases (i.e., Design, Source, Build, Analyzed, Deployed) during which an SBOM can be created or utilized.
The top part of \Cref{fig:examples} shows our taxonomy of SBOM files, highlighting in green the types we study in this work.
We refer to these files as \textit{policy-driven SBOM}---i.e., SBOM created to fulfill a security or legal policy. 
The bottom part of \Cref{fig:examples} shows an example of each type of SBOM what we consider (on the right) and that we \textit{do not} consider in this study (e.g., test data on the left).

\subsection{Related Work}
To the best of our knowledge, only a few studies focus, entirely or partially, on policy-driven SBOMs.
Torres-Arias et al.~\cite{Torres-Arias2023a} studied SBOMs obtained from two sources, ``In-The-Wild''---i.e., from open-source repositories and ``In-The-Lab''---i.e., which they created ad-hoc from a repository or a Docker image. 
As in this study, also Torres-Arias et al. used SourceGraph to collect their In-The-Wild dataset, resulting in 53 SBOM of this type. 
The authors used their dataset to benchmark two tools to assess the quality of SBOMs.
They also present a research agenda about SBOM quality aspects not captured by automated solutions, such as coverage.

Nocera et al.~\cite{Nocera2023a} studies the adoption of SBOM in open-source projects hosted on GitHub.
Conversely to our approach, the authors identified repositories containing SBOM files by searching a repository dependency graph for known tools or libraries used to generate SBOM. 
They identified 186 repositories providing SBOM and studied their evolution.

Interlynk\footnote{\url{https://www.interlynk.io}}---a commercial organization---provides tools for automated SBOM compliance, including \texttt{sbomqs} used to calculate the SBOM quality rating in this study. 
They also maintain a public database of SBOM files found In-The-Wild\footnote{\url{https://github.com/interlynk-io/sbomdb}}. 
However, these datasets include limited information---e.g., project name/URL, the tool used to generate them, format, and provenance.
Our study expanded the collected information to facilitate further studies (e.g., related to licenses or vulnerability management).

Recently, Soeiro et al.~\cite{soeiro2025wildsbomslargescaledataset} extracted 78,612 unique SBOMs by mining the Software Heritage Archive, which spans over 1,782 software forges\footnote{https://www.softwareheritage.org}. 
The authors provide information about SBOM quality, measured using the \texttt{sbomqs} tool, and track their evolution over time. 
However, they do not provide information related to vulnerabilities and licensing.

Similarly to our study, O'Donoghue et al.~\cite{o2024assessing} assess the vulnerabilities associated with SBOM files. 
To that end, they use a synthetic dataset---i.e., obtained from Interlynk---and two tools, Trivy and Grype, to obtain data related to the vulnerabilities of the dependencies declared in the SBOM (e.g., CVE, CVSS score).
They found approximately 350,000 vulnerabilities, of which approximately 10,000 were critical.
Besides basing our analysis on policy-driven SBOMs only, we use a different tool, osv-scanner, to obtain vulnerabilities.

\section{Methodology}
\label{sec:methodology}
\begin{figure*}
\centering
\includegraphics[width=.85\textwidth, trim={0cm 0cm 0.5cm 0cm},clip]{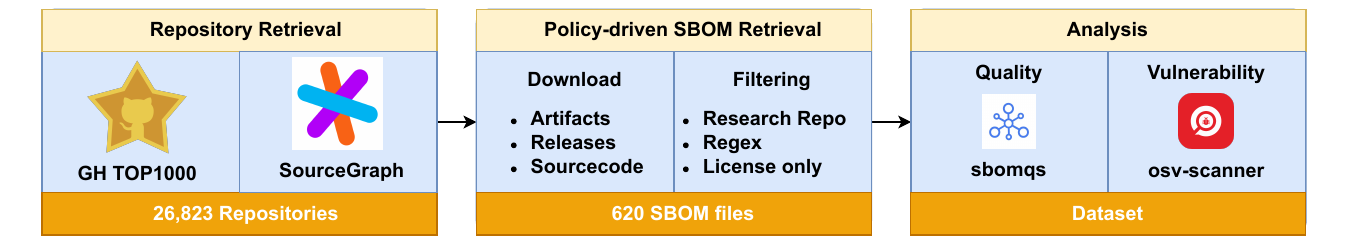}
\caption{Overview of the steps for gathering and analyzing SBOM files.}
\label{fig:flowchart}
\end{figure*}

Based on our goal, we devised the following Research Questions (RQ) to understand the current status of policy-driven SBOM. 
\begin{description}[leftmargin=*]
    \item[RQ1.]\textit{What is currently the prevalence of policy-driven SBOM in open-source projects?} Our aim is to determine how often SBOMs are created as part of a policy (e.g., for risk assessment) compared to other purposes (e.g., providing test fixtures).
    We provide a snapshot of the current state of practice in the light of upcoming legislation, such as EU CRA. 
    \item[RQ2.]\textit{What are the structural properties of policy-driven SBOM?} We investigate the formats used to create SBOMs, their quality, and components.
    \item[RQ3.] \textit{What is the current state of policy-driven SBOM regarding vulnerability management and licensing?} Policy-driven SBOM support policies related to vulnerability management and licensing obligations. This RQ focuses on information relevant to these policies, such as classes and severity of vulnerabilities and types of licenses.
 
\end{description}
To answer the RQs, we performed an \textit{explorative} mining software repository study~\cite{hassan2006mining}.
We collect and analyze the contents and quality of \sbomfiles{} SBOM files mined from GitHub.  
We associate known vulnerabilities from 28 databases with the dependencies listed in the SBOM to provide insights into the risks these dependencies pose.
\Cref{fig:flowchart} summarizes the steps we followed in our research.

\subsection{Repository Retrieval}
\label{sec:gathering}
We retrieved repositories in two steps.
First, we retrieve the most starred repositories from GitHub for each of the most popular programming languages as reported by TIOBE\footnote{\url{https://www.tiobe.com/tiobe-index/}} and GitHub Advanced Search\footnote{\url{https://github.com/search/advanced}}, as per April, 2025.

For this search, we set an initial threshold of 100 stars.
If we could not retrieve at least 1000 repositories for a given language, we lowered the threshold until we reached the lowest threshold of 35 stars.
Accordingly, we retrieved \totalGHrepositories{} repositories from GitHub and searched them for policy-driven SBOM files.

For the second phase of the search, we leveraged the SourceGraph\footnote{\url{https://sourcegraph.com/search}} (version 6.2.0) code search engine, looking for SBOM across all of the indexed repositories ($>$ 1 million).
To that end, we created regexes specific to the two most popular SBOM formats (i.e., SPDX and CycloneDX) based on their required attributes (i.e., containing the string \texttt{SPDXID} for SPDX and \texttt{bomFormat} for CycloneDX) and file extensions (e.g., \texttt{.yaml}, \texttt{.spdx}, \texttt{.xml}, and \texttt{.json}).
We applied 19 regex patterns\footnote{The regexes are available in the replication package} retrieving \totalGHandSGrepositories{} repositories---\totalGHrepositories{} directly obtained from GitHub and additional \deduplicatedleftSGrepos{} from SourceGraph after removing duplicates.
To retrieve policy-driven SBOM, we download data from the GitHub Releases,  Artifact pages, and source code. 
Specifically, considering the latest release, we downloaded (and extracted) every archive and file available and later applied the same regex described above.
From a repository release, we downloaded all artifacts archives smaller than 100MB (due to API limitation). However, in cases where any of the archives did not meet the threshold (i.e., $>$ 100MB), we only downloaded the largest among them instead.
Then, we checked that these archives related to the latest available commit in the repository's main branch and searched for SBOM files within them using the same regex. 
Moreover, we searched workflow artifacts (e.g., GitHub Actions) in every repository but were not able to retrieve any additional SBOM files.

\subsection{Policy-driven SBOM retrieval}
We are interested in studying SBOM files used to implement risk management or legal policies (e.g., vulnerability management or license compliance).
We started by manually analyzing repositories containing SBOMs to identify---based on their description and \textit{README} instructions---the ones belonging to research projects (i.e., \textit{In-the-Lab}), such as BOM shelter~\cite{Torres-Arias2023a}.
In this step, we filtered out 25 repositories. 
We then manually analyze the fully-qualified name (FQN) of SBOM files in each of the remaining repositories, looking for strings that could identify irrelevant categories---such as \textit{Test Data}, \textit{3rd Party}, and \textit{Self-License}---and adding such strings to our regex. 
We continued this process iteratively until we could not find any more SBOM files to exclude.
To that end, we applied 29 additional regex rules.
\Cref{tab:sbom_queries} reports the strings used to exclude repositories containing SBOM files.
To verify that these are \textit{policy-driven}---i.e., none belonged to the excluded categories, we manually checked a representative sample.
Given that there is a small variability across files, and for a 95\% confidence level and 5\% margin of error, a representative sample is at least $n$=96~\cite{taherdoost2017determining}.
Therefore, from the remaining \sbomfiles{} files, we randomly sampled 100 SBOM. 
In total, we retained \sbomfiles{} policy-driven SBOMs.
\Cref{tbl:repo_stats} reports descriptive statistics for repositories containing SBOMs.

\begin{table}[h]
    \centering
    \caption{Exclusion criteria for repositories with SBOM. For \textit{Test Data} and \textit{3rd Party} regex applied to filename, for \textit{Self-License} applied to SBOM content.}
    \begin{tabularx}{\columnwidth}{>{\raggedright\arraybackslash}m{0.2\columnwidth}>{\raggedright\arraybackslash}m{0.3\columnwidth}>{\raggedright\arraybackslash}m{0.4\columnwidth}}
        \toprule
        \textbf{Exclusion criteria} & \textbf{Regex} & \textbf{Example excluded repository} \\ \toprule
        Test Data & example, expect, test, demo, sample, results & CycloneDX/cyclonedx-dotnet-library, cybeats/sbomgen  \\ \midrule
        3rd Party & bundled, fixture, contrib*, dependenc*, lib*, modules, package* & apache/airflow-site, mercedes-benz/sechub\\  \midrule
        Self-License & \textit{contains a reference to a license but has empty dependencies} & allusive-dev/compfy \\
        \bottomrule
    \end{tabularx}
    \label{tab:sbom_queries}
\end{table}

\begin{table}[ht]
\centering
\caption{Summary statistics for repositories containing policy-driven SBOMs.}
\sisetup{
    table-number-alignment = right,
    table-figures-integer = 7,
    table-figures-decimal = 2,
    table-figures-uncertainty = 0,
    group-separator = {,},
    group-minimum-digits = 4
}
\begin{tabular}{l S S S}
\toprule
 & {\textbf{Stars}} & {\textbf{Contributors}} & {\textbf{Commits}} \\
\toprule
Median   & 303        & 29           & 1304       \\
\rowcolor{gray!25}Mean   & 4822.21  & 212.93  & 12369.28  \\
Std. dev. & 11235.29     & 884.06  & 83572.10  \\
\rowcolor{gray!25}Min   & 6            & 0            & 2            \\
Max   & 95443        & 13551        & 1133064      \\
\bottomrule
\end{tabular}
\label{tbl:repo_stats}
\end{table}

\subsection{Analysis}
We use the \texttt{sbomqs} tool\footnote{\url{https://github.com/interlynk-io/sbomqs}} (version 1.0.4) to calculate the quality score of each policy-driven SBOM file.
The tool calculates the quality score, between 0 and 10, based on the completeness of the SBOM structural properties (e.g., elements used based on the format specifications), its semantics (e.g., proper versioning for licenses), and the recommendations from the National Telecommunications and Information Administration agency of the USA government (NTIA)---i.e., inclusion of information regarding supplier name, component name, component version, other identifiers for the same component, dependency relationship, author, and timestamp.
We selected this tool as it incorporates several SBOM quality guidelines and supports customizable quality checklists~\cite{Mirakhorli24}.

We scan the SBOM files for vulnerabilities reported in the Google Open-Source Vulnerabilities Database using its associated tool \texttt{osv-scanner}\footnote{\url{https://osv.dev}} (version v2.0.1).
The tool uses 28 advisory databases (e.g., GitHub Advisory Database, PyPi Advisory Database) covering 21 software ecosystems (e.g., Maven, crates.io) to identify known vulnerabilities in the dependencies reported in the SBOM files.
The scan results include the Common Vulnerabilities and Exposures (CVE) identifier, the Common Weakness Enumeration (CWE) identifier, the Common Vulnerability Scoring System (CVSS) score, and the severity of the vulnerability, as well as a natural language description.
Despite other tools reported in the literature (e.g.,~\cite{o2024assessing,YU2024}), such as Trivy or Grype, use the same databases, we selected osv-scanner as it provides a standardized format---called osv-schema---which simplifies parsing and analyzing its results.
We created a dataset including information from SBOMs, their quality, and associated vulnerabilities.

\section{Results}
\label{sec:results}
This section presents the results of applying our methodology to identify policy-driven SBOMs and analyzing their contents through descriptive statistics.
\subsection{Prevalence of policy-driven SBOM (RQ1)} 
Of the \totalGHandSGrepositories{} processed repositories, \uniquerepositories{} repositories contained policy-driven SBOM files---i.e., 0.56\%. 
The most used format is SPDX, with 336 files, whereas 284 are in CycloneDX format. 
For CycloneDX, the majority uses the latest version 1.6.
Conversely, most SBOM in SPDX format use version 2.3---i.e., the second most recent version as per April 2025.
For SPDX, its latest version, 3.0, is not used by any policy-driven SBOM.  
Regarding representation, CycloneDX uses JSON and XML, whereas SPDX uses JSON, YAML and their own tag-value format (i.e., \texttt{.spdx}).
In particular, across formats, 542 SBOM use JSON, 23 use XML, 2 YAML, and 53 the SPDX tag-value.
\Cref{tbl:formats-wide} shows a breakdown of SBOM formats according to their versions.
Moreover, we found that most SBOM files, 313, belong to repositories using the Go language.
SBOM were found in repositories using popular languages, such as Java, with 41, C with 37, Rust with 25, and Python with 21 files.
\begin{table}[ht]
  \centering
  \caption{Format and version of retrieved SBOM files.}
  \label{tbl:formats-wide}
  \setlength{\arrayrulewidth}{1pt}
  \begin{tabular}{@{}l r r| l r r@{}}
    \toprule
    \textbf{Format} & \textbf{Version} & \textbf{No.\ SBOM}
     & \textbf{Format} & \textbf{Version} & \textbf{No.\ SBOM} \\
    \midrule
    \multirow{7}{*}{CycloneDX}
      & 1.6\cellcolor{gray!25} &  97\cellcolor{gray!25}
      & \multirow{4}{*}{SPDX}
      & 3.0\cellcolor{gray!25} &   0\cellcolor{gray!25}  \\
      & 1.5                      &  51
      &                           & 2.3                     & 278  \\
      & 1.4\cellcolor{gray!25}   &  71\cellcolor{gray!25}
      &                           & 2.2\cellcolor{gray!25}   &  54\cellcolor{gray!25}  \\
      & 1.3                      &  59
      &                           & 2.1                     &   4  \\
      & 1.2\cellcolor{gray!25}   &   5\cellcolor{gray!25}
      &                           &                         &      \\
      & 1.1                      &   0
      &                           &                         &      \\
      & 1.0\cellcolor{gray!25}   &   1\cellcolor{gray!25}
      &                           &                         &      \\
    \bottomrule
  \end{tabular}
\end{table}
\begin{rq:answer}[frametitle=Answer to RQ1]
Only 0.56\% of top open-source repositories use \textit{policy-driven} SBOM. The majority of SBOM do not use the latest version available for their format (SPDX or CycloneDX). 
The Go community is the most aware regarding the use of SBOM.
\end{rq:answer}

\subsection{Properties of policy-driven SBOM (RQ2)}
\begin{figure}
    \centering
    \includegraphics[width=0.8\linewidth]{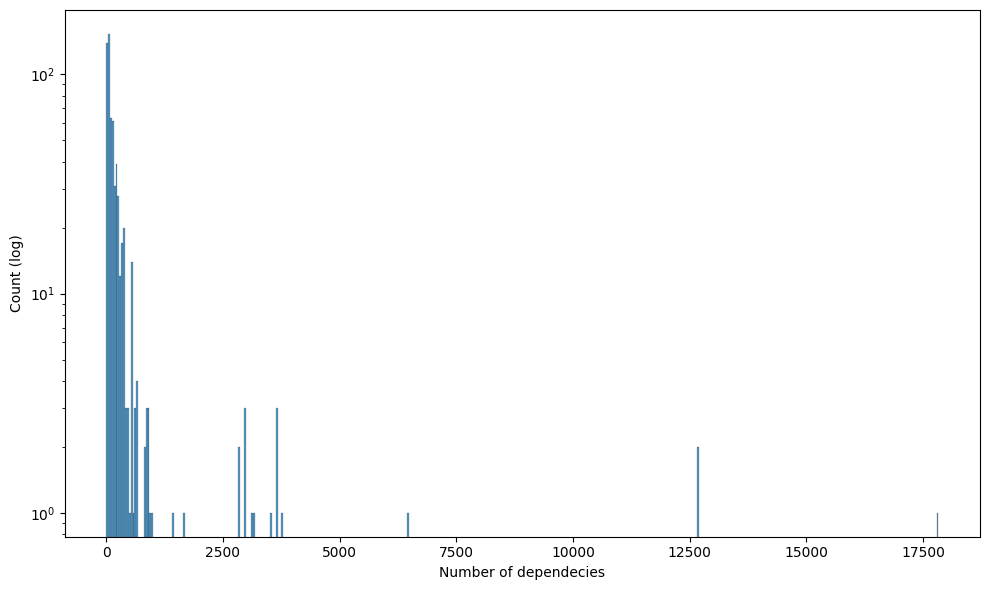}
    \caption{Distribution of dependencies declared in SBOM.}
    \label{fig:deps}
\end{figure}

In this section, we focus on properties of policy-driven SBOM.
In total, the SBOMs we collected report \dependencies{} dependencies, of which \uniquedependencies{} are unique.
On average, a policy-driven SBOM of an open-source project contains 291.86 dependencies, with a median of 93 and a standard deviation of 1131.23. 
Such variation (see~\Cref{fig:deps}) is due to large projects, such as \texttt{jaegertracing/jaeger} reporting 17,825 dependencies whereas several projects report only a single one.

\begin{table*}[t]
\centering
\caption{Summary of SBOM quality score according to their formats.}
\footnotesize
\begin{tabular}{lrrrrrr}
\toprule
 & \textbf{CDX (JSON)}& \textbf{CDX (XML)} & \textbf{SPDX (JSON)} & \textbf{SPDX (tag)} & \textbf{SPDX (YAML)} & \textbf{Total}\\
\toprule
\rowcolor{gray!25}Median & 6.95 & 8.17 & 7.36 & 6.09 & 8.17 & 7.32 \\
Mean & 6.99 & 7.61 & 7.16 & 6.65 & 8.17 & 7.07\\
\rowcolor{gray!25}Std & 0.96 & 1.33 & 0.54 & 1.38 & 0.00&0.88  \\
Min & 3.04 & 4.35 & 4.13 & 4.35 & 8.17 &3.04\\
\rowcolor{gray!25}Max & 9.38 & 8.68 & 8.18 & 9.42 & 8.17 &9.42\\
\bottomrule
\end{tabular}
\label{tbl:quality_format}
\end{table*}
The average quality of the SBOMs in our dataset is 7.07 (median = 7.32) with some variation (standard deviation = 0.88) across them. 
Specifically, both CycloneDX and SPDX SBOM have an average quality of 7, although the former has a slightly higher standard deviation than the latter (1.00 vs 0.76). 
The best score, 9.42, is achieved by the \texttt{intel/cve-bin-tool} project, whereas the project with the lowest SBOM quality of 3.04 is \texttt{scanoss/engine}.
\Cref{tbl:quality_format} reports the quality score breakdown according to SBOM formats and their representations.
\begin{rq:answer}[frametitle=Answer to RQ2]
Although used to document a large number of dependencies,  on average, the quality of policy-driven SBOM is good (i.e., 7 on a 0--10 scale).  
\end{rq:answer}

\subsection{Current state of policy-driven SBOM (RQ3)}
We analyze SBOMs from the perspective of the main policies they serve---i.e., vulnerability management and licensing.
\begin{table}[t]
    \centering
    \caption{Summary of vulnerabilities found in dependencies declared in SBOM files according to their severity.}
    \footnotesize
    \sisetup{
    table-number-alignment = right,
    table-figures-integer = 7,
    table-figures-decimal = 3,
    table-figures-uncertainty = 0,
    group-separator = {,},
    group-minimum-digits = 4
}
    \begin{tabular}{l S S S S S}
        \toprule
        & \textbf{Critical} & \textbf{High} & \textbf{Medium} & \textbf{Low} & \textbf{Total}\\
        \toprule
        \rowcolor{gray!25}Mean & 0.55 & 2.65 & 3.89 & 0.51 & 7.61  \\
        Std & 3.17 & 14.64 & 20.45 & 2.34 & 39.83\\
        \rowcolor{gray!25}Min & 0& 0 & 0 & 0 & 0\\
        Max & 56& 232& 296& 37& 296\\
        \bottomrule
    \end{tabular}
    \label{tab:severity}
\end{table}

\textit{Vulnerability management.} A large part (i.e., 39.15\%) of the dependencies reported in SBOMs did not contain any vulnerabilities.
On average, we observed 7.61 vulnerabilities per SBOM (standard deviation = 39.83). 
Projects, such as \texttt{aboutcode-org/dejacode}, show a large number of vulnerabilities (n = 564), explaining the observed variability.
On average, there are more \textit{high} and \textit{medium} severity vulnerabilities than \textit{critical}, while \textit{low} ones appear less often.
\Cref{tab:severity} reports vulnerabilities information related to SBOM files.

\begin{figure}[!b]
  \centering
  \begin{subfigure}[t]{0.49\textwidth}
    \centering
    \includegraphics[trim={0 0.8cm 0 0},clip,width=1\linewidth]{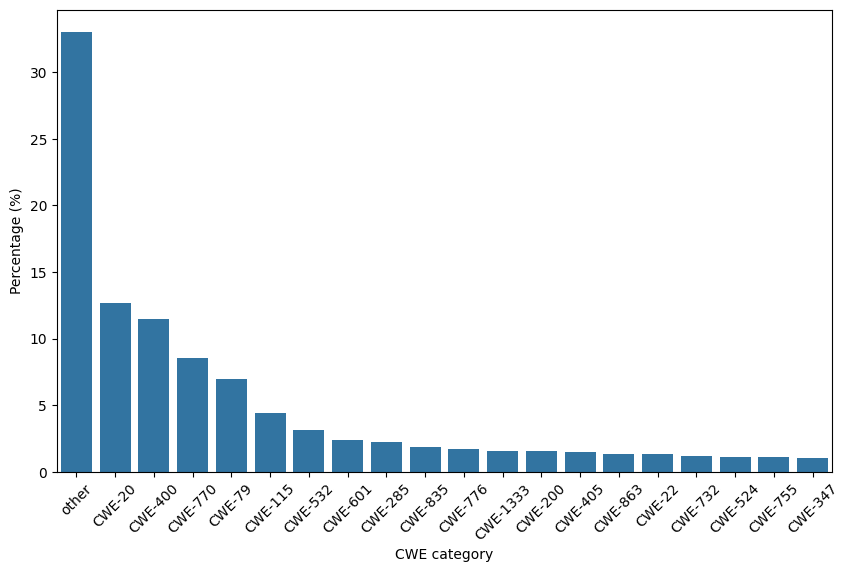}%
    \caption{CWE distribution in SBOM files.}
    \label{fig:cwe}%
  \end{subfigure}
  \begin{subfigure}[t]{0.49\textwidth}
    \centering
    \includegraphics[trim={0 0.8cm 0 0},clip,width=1\linewidth]{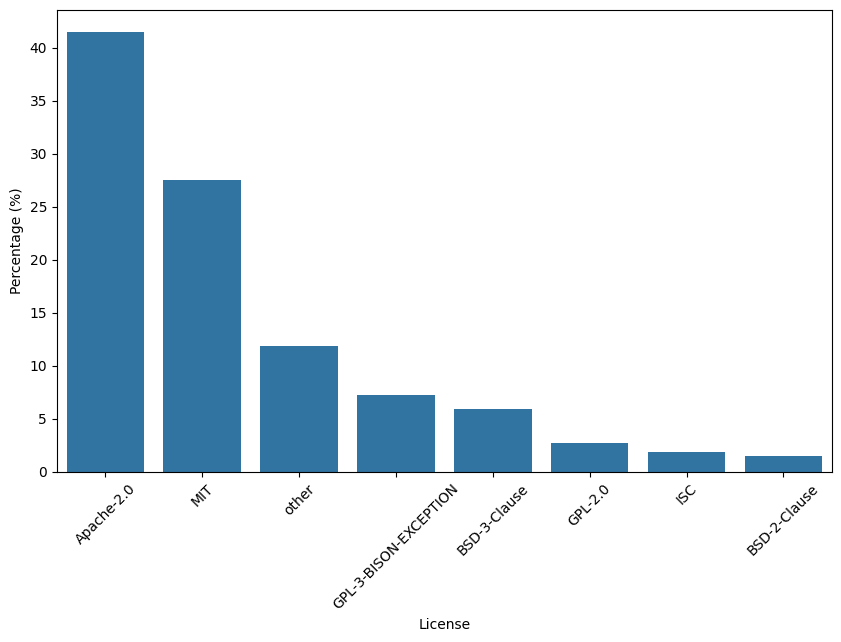}%
    \caption{License distribution in SBOM files.}
    \label{fig:license}%
  \end{subfigure}
  \caption{SBOM vulnerability and licensing information.}
  \label{fig:combined}
\end{figure}

In total, we identified 19,225 CVEs, of which 2,202 tracking unique vulnerabilities.
The most common vulnerability is tracked as CVE-2025-22872, which impacts 824 dependencies.
The oldest vulnerabilities is from 2012, and is tracked as CVE-2012-0805.
All the Top-10 most common vulnerabilities (see~\Cref{tbl:cves}) were published between 2020 and 2025. 
Regarding categories, the vulnerabilities found in the dependencies of the collected SBOM files cover 223 unique CWE (8,726 in total).
On average, the vulnerabilities are associated with three CWE categories (standard deviation = 4.85, median = 2, max = 56).  
The most vulnerabilities fall under CWE-20 related  to weak input validation which can lead to other attack vectors, such as different types of injections. 
Furthermore, denial of service (e.g., through computing resource consumption), such as CWE-400, CWE-1333, and CWE-770 are the second most common types of vulnerability affecting the dependencies declared in policy-driven SBOMs. 
\Cref{fig:cwe} reports the CWE of with vulnerabilities observed in policy-driven SBOMs. 

\textit{Licensing.} The collected SBOMs report a total of 384 unique licenses.
Each SBOM file reports, on average, 107 licenses (median = 1) per project, with a  standard deviation of 370.
This is due to projects containing a high number of licenses, such as \texttt{apache/camel-quarkus} reporting 3757 licenses.

Interestingly, 139 SBOMs (22.41\%) do not report any license (83 in CycloneDX and 56 in SPDX format).
Apache and MIT are the most utilized licenses, representing approximately 70\% of the total.
\Cref{fig:license} reports the occurrences of different licenses in policy-driven SBOMs. 

\begin{rq:answer}[frametitle=Answer to RQ3]
Although the majority of dependencies reported in SBOM do not contain vulnerabilities, medium to high-severity ones are still common and unpatched.
Moreover, there are hundreds license reported, although 22\% of policy-driven SBOM do not report any licensing information.
Permissive licenses (e.g., Apache 2.0 and MIT) are the most common.
\end{rq:answer}

\section{Discussion}
\label{sec:directions}
\begin{table}[t]
\centering
\caption{Top-5 CVEs by occurrence found in dependencies declared in SBOMs.}
\footnotesize
\begin{tabular}{lrr}
\toprule
 \textbf{CVE ID}& \textbf{Deps. impacted} & \textbf{Percentage} \\ 
 \toprule
     \rowcolor{gray!25}
CVE-2025-22872 & 824 & 4.28 \\
CVE-2025-22870 & 766 & 3.98 \\
\rowcolor{gray!25}
CVE-2025-22869 & 510 & 2.65 \\
CVE-2024-45337 & 370 & 1.92 \\
\rowcolor{gray!25}
CVE-2025-22871 & 363 & 1.88 \\
     \bottomrule
\end{tabular}
\label{tbl:cves}
\end{table}
This section shows the limitations of our results and discusses their implications. 

\subsection{Limitations}
\label{sec:limitations}
Although we followed similar existing studies in the area of SBOM and supply chain security (e.g.,~\cite{YU2024,Rabbi2024,Nocera2023a,Bi2024}), our limitation is that we used GitHub as the only source for SBOM.

SourceGraph only searches within the contents of files smaller than 1MB.
Other limitations that could impact our results deal with the SourceGraph indexing strategy, which limits our search to the main branch of a repository.
Moreover, we could not search repositories larger than 10GB as they are not indexed.
Our results are based on the top-1000 repositories for each language.
Accordingly, we acknowledge that the results could differ for SBOM files belonging to projects utilizing less popular programming languages or simply less popular projects.
However, our results are based on an initial set of \totalGHandSGrepositories{} repositories which we believe to be representative.

Another limitation of our study is the regex used to search for policy-driven SBOM files. 
There is a chance that our regexes filtered out actual policy-driven SBOM (i.e., false negatives).
Moreover, we could have included irrelevant SBOM files due to their non-standard template, such as mis-using keywords used in our regexes (i.e., false positives). 
However, through manual analysis, we detected non-policy-driven SBOMs and iteratively improved the regex used for filtering them out. 
A manual inspection of a significant sample of the final dataset showed no false positives.
We limited our search to SPDX and CycloneDX, as these are the two most widespread formats~\cite{Torres-Arias2023a,balliu2023,Nocera2023a}.

Finally, we acknowledge a limitation due to the SBOM tools we used. 
To calculate the SBOM scores, we have used a single tool, although we acknowledge that using different tools (e.g., OWASP SCVS, Ebay's SBOM scorecard) could yield different results~\cite{Mirakhorli24}.
The tool could not parse 66\% of SBOM files in SPDX \texttt{tag-value} format across versions due to missing fields and invalid relationship type.
We tried to address this by converting the tag-value format to JSON, using the official tool provided by SPDX, to no avail.
The \texttt{osv-scanner} tool failed to obtain information from the advisory databases for 22\% of the SBOMs in our dataset.
Similar issues are reported in previous research (e.g.,~\cite{Mirakhorli24}) showing that SBOM tooling is currently immature.
\subsection{Implications}
As reported in this paper and previous studies (e.g.,~\cite{Torres-Arias2023a}), there are very few publicly available SBOMs compared to the number of open-source projects.
Further, the results of RQ1 show that only a small percentage of repositories contain SBOM created and shared as part of a policy.   
By empirically investigating the current state of policy-driven SBOM and their contents, we aim to improve SBOM research, specifically the study of SBOM usage by software providers.
\takeaway{1}{Researchers investigating SBOM mined from GitHub should consider whether the SBOM they are investigating are \textit{policy-driven} or not, or at least consider whether they are analyzing In-the-Wild or In-the-Lab SBOM.}
For researchers, further analysis of our dataset will serve as the first step towards answering fundamental questions, such as ``\textit{What is actually documented in an SBOM?}'' and ``\textit{How does SBOM support risk assessment in practice?}''

Also in RQ1, we show that only 34\% of CycloneDX SBOM are using the latest standard version, whereas 20\% are still using a version from January 2022.
For SPDX, although the latest major version of the standard is not in use yet In-The-Wild at the time of this study, we show that 16\% of SBOM use a version (v2.2) from May 2020.
\takeaway{2}{Software vendors should consider updating their SBOM to the latest available version of the most popular formats to avoid compatibility issues in the future. Software integrators operating down the supply chain should be aware of older SBOM format version limitations, such as lack of attestation.}

The results of RQ2 show that the policy-driven SBOM are of good quality and can be used as an initial benchmark for In-the-Wild evaluation of tools that use SBOM as part of their input, such as vulnerability scanners (e.g.,~\cite{Imtiaz.2021,CAO23}). 
Such evaluation can inform practitioners in selecting an appropriate tool for downstream tasks (e.g., scanning for licenses). 
Similarly to previous work (i.e.,~\cite{Torres-Arias2023a}), we contribute to the research area of SBOM quality. 
The analysis of policy-driven SBOMs can guide the development of an empirically-based definition of what is an SBOM of high quality and what is not in different contexts (e.g., at different stages of a software supply chain) and for different use cases~\cite{Mirakhorli24}. 

Given the current legislation in the US and the upcoming one in Europe, SBOM will be integrated into the development lifecycle to help identify known vulnerabilities in the listed dependencies. 
In our study (RQ3), we show that high-severity vulnerabilities affecting dependencies are left unpatched for several months.
Connecting vulnerability information to an SBOM can support suppliers and integrators to quickly assess the security posture of a software product, leading to better risk mitigation strategies for the users and downstream actors in the supply chain.
The security community is moving in this direction by developing the Vulnerability Exploitability eXchange (VEX) format~\cite{stalnaker2024boms}.
\takeaway{3}{Open source project maintainers as well as software vendors using open source should be aware that upcoming regulations, such as EU CRA, will enforce strict time limits for patching vulnerabilities and updating vulnerable dependencies.}
Further research needs to deal with how to use the information contained in SBOMs, aligning them with vulnerability databases, and communicating these vulnerabilities to development teams so they can act upon them and make informed decision when using a third-party library.
When investing RQ3, we show that there are several unique licenses (107 on average) appearing in the supply chain of an open-source project. 
Managing such variety can cause issues, from a legal standpoint, for detecting and resolving conflicts between licenses.
Tools developed for analyzing licenses (e.g.,~\cite{xu2023lidetector}) can use the dataset provided in this research to created benchmarks of the current state of open-source licenses for components documented in SBOM files.
The large number of licenses in dependencies should be a warning sign for software vendors.
\takeaway{4}{Researchers should investigate how SBOMs are explicitly used to support licensing of software dependencies (e.g., resolving incompatibility issues). Software vendors should be aware that SBOM can help them review possible licensing issues related to the software they provide.}
Our analysis of both vulnerabilities and licenses contained in policy-driven SBOM can guide researchers to develop approaches for automatic recommendation of dependency replacements---e.g., when a dependency declared in an SBOM contains vulnerabilities above a certain risk threshold or the dependency is incompatible with the product licensing scheme. 

\section{Conclusion}
\label{sec:conclusion}
This study analyzed \sbomfiles{} policy-driven SBOM files mined from open-source projects.
The results show that only 0.56\% of repositories contain SBOM that serve a policy, such as vulnerability risk assessment or licenses management.
Structural properties of policy-driven SBOMs show a good level of quality.
We show that policy-driven SBOM in GitHub still contain high-severity unpatched CVEs, which could result in a violation of the upcoming EU CRA.
The large number of licenses of dependencies we found can cause issues when these are not properly managed (e.g., license conflicts). 
We emphasize the importance of investigating policy-driven SBOMs to improve supply chain security.

Future research will focus on the time-related aspects of \textit{policy-driven} SBOMs. 
This includes understanding the time frame for their inclusion in a project and determining whether specific formats and versions become prevalent over time.
Moreover, we will investigate the motivation for the low adoption of SBOM, repository maintainers policy to create and maintain them.

\bibliographystyle{unsrt}
\bibliography{sbom}

\begin{thebibliography}{10}

\bibitem{feng2022defense}
Sylvia Feng and Muharman Lubis.
\newblock Defense-in-depth security strategy in log4j vulnerability analysis.
\newblock In {\em 2022 International Conference Advancement in Data Science,
  E-learning and Information Systems (ICADEIS)}, pages 01--04. IEEE, 2022.

\bibitem{wolff2021navigating}
Evan~D Wolff, KM~Growley, MG~Gruden, et~al.
\newblock Navigating the solarwinds supply chain attack.
\newblock {\em The Procurement Lawyer}, 56(2), 2021.

\bibitem{arvanitis2022systematic}
Iosif Arvanitis, Grigoris Ntousakis, Sotiris Ioannidis, and Nikos Vasilakis.
\newblock A systematic analysis of the event-stream incident.
\newblock In {\em Proceedings of the 15th European Workshop on Systems
  Security}, pages 22--28, 2022.

\bibitem{everson2022log4shell}
Douglas Everson, Long Cheng, and Zhenkai Zhang.
\newblock Log4shell: Redefining the web attack surface.
\newblock In {\em Workshop on Measurements, Attacks, and Defenses for the Web
  (MADWeb) 2022}, 2022.

\bibitem{Zahan.20236zj}
Nusrat Zahan, Elizabeth Lin, Mahzabin Tamanna, William Enck, and Laurie
  Williams.
\newblock Software bills of materials are required. are we there yet?
\newblock {\em IEEE Security \& Privacy}, 21(2):82--88, 2023.

\bibitem{hendrick2022state}
Stephen Hendrick and J~Zemlin.
\newblock The state of software bill of materials (sbom) and cybersecurity
  readiness.
\newblock Technical report, The Linux Foundation, 2022.

\bibitem{kloeg2024charting}
Berend Kloeg, Aaron~Yi Ding, Sjoerd Pellegrom, and Yury Zhauniarovich.
\newblock Charting the path to sbom adoption: A business stakeholder-centric
  approach.
\newblock In {\em Proceedings of the 19th ACM Asia Conference on Computer and
  Communications Security}, pages 1770--1783, 2024.

\bibitem{stalnaker2024boms}
Trevor Stalnaker, Nathan Wintersgill, Oscar Chaparro, Massimiliano Di~Penta,
  Daniel~M German, and Denys Poshyvanyk.
\newblock Boms away! inside the minds of stakeholders: A comprehensive study of
  bills of materials for software systems.
\newblock In {\em Proceedings of the 46th IEEE/ACM International Conference on
  Software Engineering}, pages 1--13, 2024.

\bibitem{Xia2023}
Boming Xia, Tingting Bi, Zhenchang Xing, Qinghua Lu, and Liming Zhu.
\newblock An empirical study on software bill of materials: Where we stand and
  the road ahead.
\newblock {\em arXiv}, 2023.

\bibitem{Torres-Arias2023a}
Santiago Torres-Arias, Dan Geer, and John~Speed Meyers.
\newblock A viewpoint on knowing software: Bill of materials quality when you
  see it.
\newblock {\em IEEE Security \& Privacy}, 21(6):50--54, 2023.

\bibitem{Mirakhorli24}
Mehdi Mirakhorli, Derek Garcia, Schuyler Dillon, Kevin Laporte, Matthew
  Morrison, Henry Lu, Viktoria Koscinski, and Christopher Enoch.
\newblock A landscape study of open source and proprietary tools for software
  bill of materials (sbom).
\newblock {\em arXiv preprint arXiv:2402.11151}, 2024.

\bibitem{o2024assessing}
Eric O’Donoghue, Ann~Marie Reinhold, and Clemente Izurieta.
\newblock Assessing security risks of software supply chains using software
  bill of materials.
\newblock In {\em 2nd International Workshop on Mining Software Repositories
  for Privacy and Security, MSR4P\&S,(SANER 2024), Rovaniemi, Finland}, 2024.

\bibitem{balliu2023}
Musard Balliu, Benoit Baudry, Sofia Bobadilla, Mathias Ekstedt, Martin
  Monperrus, Javier Ron, Aman Sharma, Gabriel Skoglund, C{\'e}sar Soto-Valero,
  and Martin Wittlinger.
\newblock Challenges of producing software bill of materials for java.
\newblock {\em IEEE Security \& Privacy}, 2023.

\bibitem{YU2024}
Sheng Yu, Wei Song, Xunchao Hu, and Heng Yin.
\newblock On the correctness of metadata-based sbom generation: A differential
  analysis approach.
\newblock In {\em 2024 54th Annual IEEE/IFIP International Conference on
  Dependable Systems and Networks (DSN)}, pages 29--36. IEEE, 2024.

\bibitem{Rabbi2024}
Md~Fazle Rabbi, Arifa~Islam Champa, Costain Nachuma, and Minhaz~Fahim Zibran.
\newblock Sbom generation tools under microscope: A focus on the npm ecosystem.
\newblock In {\em Proceedings of the 39th ACM/SIGAPP Symposium on Applied
  Computing}, pages 1233--1241, 2024.

\bibitem{vendome2017license}
Christopher Vendome, Gabriele Bavota, Massimiliano~Di Penta, Mario
  Linares-V{\'a}squez, Daniel German, and Denys Poshyvanyk.
\newblock License usage and changes: a large-scale study on github.
\newblock {\em Empirical Software Engineering}, 22:1537--1577, 2017.

\bibitem{CISA}
Kate Stewart and Melissa Rhodes.
\newblock Types of software bill of materials (sbom).
\newblock
  \url{https://web.archive.org/web/20240907164647/https://www.cisa.gov/sites/default/files/2023-04/sbom-types-document-508c.pdf}.

\bibitem{Nocera2023a}
Sabato Nocera, Simone Romano, Massimiliano Di~Penta, Rita Francese, and
  Giuseppe Scanniello.
\newblock Software bill of materials adoption: a mining study from github.
\newblock In {\em 2023 IEEE International Conference on Software Maintenance
  and Evolution (ICSME)}, pages 39--49. IEEE, 2023.

\bibitem{soeiro2025wildsbomslargescaledataset}
Luıs Soeiro, Thomas Robert, and Stefano Zacchiroli.
\newblock Wild sboms: a large-scale dataset of software bills of materials from
  public code, 2025.

\bibitem{hassan2006mining}
Ahmed~E Hassan.
\newblock Mining software repositories to assist developers and support
  managers.
\newblock In {\em 2006 22nd IEEE International Conference on Software
  Maintenance}, pages 339--342. IEEE, 2006.

\bibitem{taherdoost2017determining}
Hamed Taherdoost.
\newblock Determining sample size; how to calculate survey sample size.
\newblock {\em International Journal of Economics and Management Systems}, 2,
  2017.

\bibitem{Bi2024}
Tingting Bi, Boming Xia, Zhenchang Xing, Qinghua Lu, and Liming Zhu.
\newblock On the way to {SBOM}s: Investigating design issues and solutions in
  practice.
\newblock {\em ACM Transactions on Software Engineering and Methodology},
  page~0, 2024.

\bibitem{Imtiaz.2021}
Nasif Imtiaz and Laurie Williams.
\newblock Memory error detection in security testing.
\newblock {\em arXiv}, 2021.

\bibitem{CAO23}
Dinis~Barroqueiro Cruz, João~Rafael Almeida, and José~Luís Oliveira.
\newblock Open source solutions for vulnerability assessment: A comparative
  analysis.
\newblock {\em IEEE Access}, 11:100234--100255, 2023.

\bibitem{xu2023lidetector}
Sihan Xu, Ya~Gao, Lingling Fan, Zheli Liu, Yang Liu, and Hua Ji.
\newblock Lidetector: License incompatibility detection for open source
  software.
\newblock {\em ACM Transactions on Software Engineering and Methodology},
  32(1):1--28, 2023.

\end{thebibliography}
\end{document}